\documentclass[12pt,superscriptaddress,letter]{revtex4}%
\usepackage{amssymb}
\usepackage{color}
\usepackage{graphicx}
\usepackage{dcolumn}
\usepackage{bm}
\usepackage[header,title,page,titletoc]{appendix}
\usepackage{amsmath}
\usepackage{subfigure}
\usepackage{amsfonts}%
\providecommand{\U}[1]{\protect \rule{.1in}{.1in}}
\begin{document}
\title{Topological Mid-gap States of $p_{x}+ip_{y}$ Topological Superconductor with
Vortex superlattice}
\author{Jiang Zhou}
\author{Shi-Zhu Wang}
\author{Ya-Jie Wu}
\author{Rong-Wu Li}
\author{Su-Peng Kou}
\thanks{Corresponding author}
\email{spkou@bnu.edu.cn}
\affiliation{Department of Physics, Beijing Normal University, Beijing 100875, PR China}

\begin{abstract}
In this paper, the $p_{x}+ip_{y}$ topological superconductor with vortex
superlattice is studied. We found that there exist mid-gap energy bands
induced by the vortex superlattice and the mid-gap energy bands have
nontrivial topological properties including the gapless edge states and
non-zero winding number. An topological anisotropic tight-binding Majorana
lattice model is proposed to describe the mid-gap states.

\end{abstract}
\maketitle

\section{Introduction}

The topological ordered states become active research fields in condensed
matter physics\cite{Wen1,Wen2}. The first example is integer quantum Hall
effect, of which people introduce a topological invariant (TKNN number) to
describe the topological properties\cite{DJThouless}. Recently, the
topological insulators with Z2 topological invariant are proposed and realized
in experiments\cite{CLKane,BABernevig}. Another class of topological quantum
states is the topological superconductor, of which an example is two
dimensional $p_{x}+ip_{y}$ topological superconductor\cite{NRead}. The
$p_{x}+ip_{y}$ superconductor (SC) has full bulk gap and topologically
protected Majorana edge states. In particular, for the two dimensional
$p_{x}+ip_{y}$ topological superconductor, the quantized vortex traps Majorana
fermion inside the vortex-core. So, there are $2^{N}$-fold degenerate ground
states for $2N$ vortices. The adiabatic exchanges of two vortices generate a
unitary transformations in the $2^{N}$ dimensional Hilbert space, which
implies the non-Abelian statistics for the vortices\cite{AdyStern}. These
Majorana zero modes trapped by the vortex-cores immune to any small
perturbations, and thus are proposed to do topological quantum
computation\cite{ki2,free,ge,CNayak}.

Recently, many theoretical and experimental efforts have been paid to the
detection of Majorana zero modes for the potential application
\cite{Jalicea,MCheng}. For the $p_{x}+ip_{y}$ SC, a single vortex associates a
rigorous Majorana zero energy bound state. However, when taking two vortices
nearby, the inter-vortex tunneling occurs and leads to a small energy
splitting which removes the ground state degeneracy. Once the vortex is
arranged regularly forming vortex lattice with vortex superlattice constant
being the order of superconducting coherent length, the inter-vortex tunneling
will modify the low energy band of $p_{x}+ip_{y}$ SC and induce mid-gap states.

In this paper, we focus on the $p_{x}+ip_{y}$ SC with vortex superlattice, of
which the lattice constant is about $l\sim2\xi$ , where the coherent length
can be estimated as the radius of profile of the wave function around the
vortex. Our goal is to learn the nature of the mid-gap states induced by the
vortex superlattice, then we use free Majorana lattice model to capture its
properties. The paper is organized as follows. In Sec. II, we review the
spinless $(p_{x}+ip_{y})$ SC and show that a Majorana fermion with zero energy
is trapped in the vortex core. In Sec. III, we present that the inter-vortex
tunneling leads to the energy splitting and show the band structure of the
mid-gap states of $p_{x}+ip_{y}$ SC with vortex superlattice. We also suggest
a topological anisotropic Majorana lattice model to capture the low energy
properties of the mid-gap states in this section. Finally, we conclude our
discussion in Sec. IV.

\section{The model of $p_{x}+ip_{y}$ superconductor}

In this section, firstly we review the $p_{x}+ip_{y}$ SC. We write down a
lattice Hamiltonian for spinless fermions. From the corresponding BCS mean
field theory, there exist two distinct phases, the weak pairing SC and the
strong pairing SC, which are distinguished topologically. Then, we analyze the
Bogoliubov-de Gennes (BdG) equation in the continuum limit and show that the
zero energy bound state (zero mode) is described by a exponentially localized
wave function.

\subsection{Two dimensional $p_{x}+ip_{y}$ superconductor}

For the sake of completeness, we now review the basis formalism of the
$p_{x}+ip_{y}$ SC. The simplest form exhibiting $p_{x}+ip_{y}$
superconductivity is encoded in the following lattice Hamiltonian $H$, where
\begin{align*}
H  &  =H_{0}+H_{1},\\
H_{0}  &  =-t\sum_{\mathbf{r}}\sum_{v=\mathbf{\hat{x}},\mathbf{\hat{y}}%
}(c_{\mathbf{r}+v}^{\dag}c_{\mathbf{r}}+h.c)-u\sum_{\mathbf{r}}c_{\mathbf{r}%
}^{\dag}c_{\mathbf{r}},\\
H_{1}  &  =-\frac{1}{2}\hat{\Delta}\sum_{\mathbf{r}}\{(c_{\mathbf{r}%
+\mathbf{\hat{y}}}c_{\mathbf{r}}-c_{\mathbf{r}-\mathbf{\hat{y}}}c_{\mathbf{r}%
})\\
&  -i(c_{\mathbf{r}+\mathbf{\hat{x}}}c_{\mathbf{r}}-c_{\mathbf{r}%
-\mathbf{\hat{x}}}c_{\mathbf{r}})\}+h.c.
\end{align*}
where $u$ is the chemical potential, $\hat{\Delta}$ is the electron pairing
function and $t$ is the hopping strength, respectively. In the following
parts, we set $t$ to be energy unit. The operator $c_{\mathbf{r}}^{\dagger
}/c_{\mathbf{r}}$ creates/destroys an electron on lattice site $\mathbf{r}$
and satisfies the anti-commutation statistics $\{c_{\mathbf{r}},c_{\mathbf{r}%
^{\prime}}^{\dag}\}=\delta_{\mathbf{rr}^{\prime}}$.

In terms of Nambu spinor $\Psi_{\mathbf{k}}^{\dag}=(c_{\mathbf{k}}^{\dag
},c_{-\mathbf{k}})$, the mean field Hamiltonian takes the form of
\begin{equation}
H=\frac{1}{2}\int_{\mathbf{k}}d\mathbf{k}\Psi_{\mathbf{k}}^{\dag}%
H(\mathbf{k})\Psi_{\mathbf{k}}%
\end{equation}
via the Fourier transformation into momentum, and $H(\mathbf{k})$ is a
$2\otimes2$ matrix that reads
\begin{equation}
H(k)=\left(
\begin{array}
[c]{cc}%
\xi(\mathbf{k}) & \Delta^{\ast}(\mathbf{k})\\
\Delta(\mathbf{k}) & -\xi(\mathbf{k})
\end{array}
\right)
\end{equation}
where $\xi(\mathbf{k})=-2t(\cos \mathbf{k}_{x}+\cos \mathbf{k}_{y})-u$. The
pairing $\Delta(\mathbf{k})=\sin \mathbf{k}_{x}+i\sin \mathbf{k}_{y}$ exhibits
the $p$-wave (spin-triplet) symmetry as $\Delta(\mathbf{k})=-\Delta
(-\mathbf{k})$. The Hamiltonian $H$ can be diagonalized by the Bogoliubov
transformation $\alpha_{\mathbf{k}}=u_{\mathbf{k}}c_{\mathbf{k}}%
-v_{\mathbf{k}}c_{-\mathbf{k}}^{\dag}$ so that $\{ \alpha_{\mathbf{k}}%
,\alpha_{\mathbf{k}^{\prime}}^{\dag}\}=\delta_{\mathbf{kk}^{\prime}}$. The
$\mathbf{k}$-dependent coefficients $u_{\mathbf{k}}$ and $v_{\mathbf{k}}$ can
be determined according to the requirement that the full Hamiltonian has the
diagonal form
\begin{equation}
H=\sum_{\mathbf{k}}E(\mathbf{k})\alpha_{\mathbf{k}}^{\dag}\alpha_{\mathbf{k}%
}+E_{g}%
\end{equation}
and the quasiparticle operator $\alpha_{\mathbf{k}}$ satisfies the
anti-commutation relation. The quasi-particle excitation spectrum is given by
\begin{equation}
E(\mathbf{k})=\sqrt{\xi^{2}(\mathbf{k})+|\Delta(\mathbf{k})|^{2}}.
\end{equation}

The energy gap closes at $|u|=4t$, which can be used to define a topological
quantum phase transition. For $|u|<4t$, the system is in the weak pairing
phase (topologically nontrivial phase). For $|u|>4t$, the system is in a
strong pairing phase (topological trivial phase). The quantum critical point
at $|u|=4t$ marks the phase transition between the weak paring phase and the
strong paring phase.

\subsection{The Majorana zero modes around SC vortex}

To demonstrate the fact that a vortex in $p_{x}+ip_{y}$ SC traps a Majorana
zero mode, we consider the low energy limit and make the substitution
$\xi(\mathbf{k})\approx-u(\mathbf{r})$. For a spatially slowly varying
$u(\mathbf{r})$, we consider a domain wall $u(\mathbf{r})<0$ for
$\mathbf{r}>\mathbf{r}_{0}$ and $u(\mathbf{r})>0$ for $\mathbf{r}%
<\mathbf{r}_{0}$. Since the different regions are in distinct topological
phases, one expects edge states at the interface. It's sufficient to start
with the continuum form
\begin{equation}
H_{BCS}=\int d^{2}\mathbf{r}\{-u(\mathbf{r})\psi^{\dagger}\psi+[\frac{\Delta
}{2}\psi^{\dagger}(\partial_{x}+i\partial_{y})\psi^{\dagger}+h.c]\}.
\end{equation}
The Bogliubov-de-Gennes (BdG) matrix corresponding to this equation reads (see
details in appendix)
\begin{equation}
H_{BdG}(\mathbf{r})=\left(
\begin{array}
[c]{cc}%
-u(\mathbf{r}) & \{ \Delta(\mathbf{r}),\partial_{\mathbf{r}}+i\partial
_{\theta}/r\} \\
-\{ \Delta^{\ast}(\mathbf{r}),\partial_{\mathbf{r}}-i\partial_{\theta}/r\} &
u(\mathbf{r})
\end{array}
\right)  , \label{bdg}%
\end{equation}
with anticommutator being defined as $\{a,b\}=[ab+ba]/2$. To find the wave
functions of the zero modes satisfying $H_{BdG}(\mathbf{r})\chi(\mathbf{r}%
)=E\chi(\mathbf{r})$, it's helpful to assume $\Delta(\mathbf{r})=\Delta
e^{-il\theta}$, which denotes the pairing with vorticity $l$ located at
position $\mathbf{r}$.

We solve the problem with the ansatz
\begin{equation}
\chi(\mathbf{r})=\left(
\begin{array}
[c]{c}%
e^{-i\theta/2}[f(r)+ig(r)]\\
e^{i\theta/2}[f(r)-ig(r)]
\end{array}
\right)
\end{equation}
where $f(r)$ and $g(r)$ obey%
\begin{align}
-iu(r)g-i\Delta \partial_{r}g-i\frac{\Delta(l-1)g}{2r}  &  =Ef,\nonumber \\
u(r)f-\Delta \partial_{r}f-\frac{\Delta(l-1)f}{2r}  &  =Eg.
\end{align}
Then, we find two zero modes located the boundary of the domain wall if $l=1.$
The wave functions of the zero modes are given by%
\begin{equation}
\chi(\mathbf{r})\sim e^{-\frac{1}{\Delta}\int_{\mathbf{r}_{0}}^{\mathbf{r}%
}d\mathbf{r}^{\prime}u(\mathbf{r}^{\prime})}(%
\begin{array}
[c]{c}%
e^{-i\theta/2}\\
e^{i\theta/2}%
\end{array}
).
\end{equation}
The exponentially localized wave function corresponds to Majorana fermions due
to the particle-hole symmetry.

To summarize, we have obtained the Majorana bound state (BS) with zero energy
attached to a single $p+ip$ superconducting vortex. For the multi-vortex case,
we need to take into account the inter-vortex tunneling.

\section{Mid-gap states}

In last section, we have reviewed that the $p_{x}+ip_{y}$ topological SC
supports zero mode around vortices. These vortices with topologically
protected zero mode obey non-Abelian statistics. In this section, we focus on
the mid-gap states induced by the vortex superlattice after considering the
coupling between Majorana fermions on different vortices.

To study the coupling between Majorana fermions on different vortices, we
assume that each vortex locally threads into single plaquette. Each vortex can
be regarded as the end of a string. This string is the phase branch-cut of the
pairing order parameter. Each (spinless) fermion denoting by $f$ acquires a
minus sign after moving around the vortex by crossing the phase branch-cut as
$f\rightarrow-f$. So the vortex is a really $\pi$-flux on a plaquette. This
definition provides us an effective method to numerically compute the
properties of the quantized vortex in the $p_{x}+ip_{y}$ topological SC. The
inset of Fig.1 shows two vortices with two approximate zero modes localized in
each core.

\begin{figure}[ptb]
\scalebox{0.34}{\includegraphics* [0.5in,0.0in][12in,7.2in]{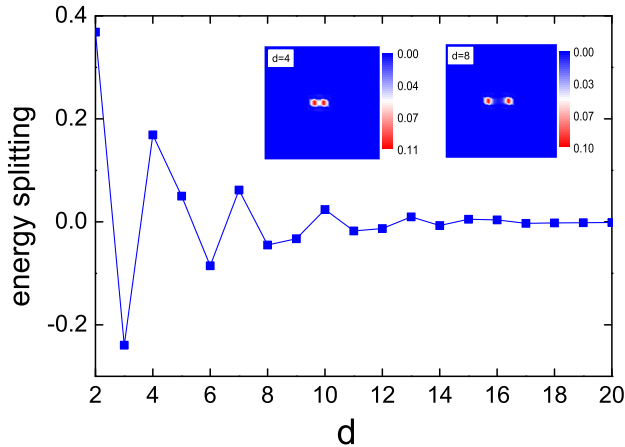}}\caption{(Color
online) The energy splitting as a function of the space distance of two
vortices. The inset shows the particle density distributions of two vortices
separated by $d=4$ and $d=8$.}%
\end{figure}

Because the inter-vortex tunneling effect leads to the energy splitting of the
two zero modes, we must take into account this tunneling effect for two
quantized vortices nearby. To see the tunneling effect clearer, we performed
numerical simulations and the results are shown in Fig.1. Generally, the
energy splitting due to inter-vortex tunneling is determined by the overlap of
the wave functions of two vortices. One can see that the energy splitting
between two nearby vortices (the distance $d$ between two vortices nearby is
smaller than the coherent length $\xi$) exhibits a relatively large value. On
the other hand, for two well separated vortices (the distance $d$ between two
vortices nearby is larger than the coherent length $\xi$), the energy
splitting from the inter-vortex tunneling can be ignored. In Fig.1, we have
chosen $\Delta=-1.0t$ to do our calculations.

\subsection{Mid-gap energy bands and its edge states induced by vortex
superlattice}

\begin{figure}[ptb]
\scalebox{0.34}{\includegraphics* [0.5in,0.0in][12in,7.2in]{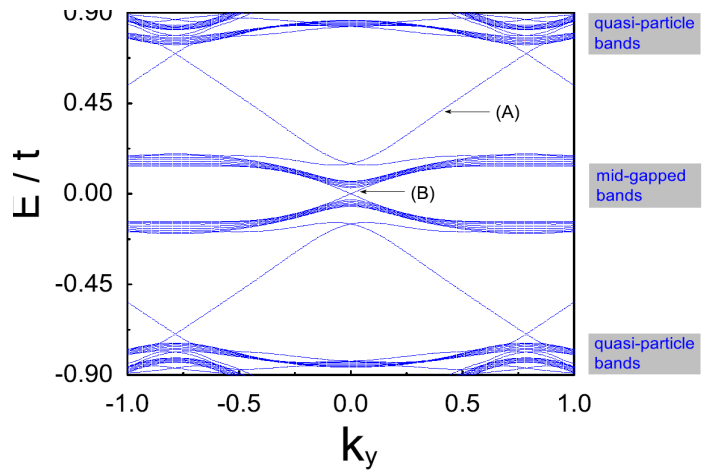}}\caption{(Color
online) The band structure of the topological superconductor with a vortex
superlattice on a cylindrical geometry. (A): the edge state of the parent
topological superconductor; (B): the edge state of the mid-gap
states induced by the vortex superlattice.}%
\end{figure}

\begin{figure}[ptb]
\scalebox{0.4}{\includegraphics* [width=1.1\textwidth]{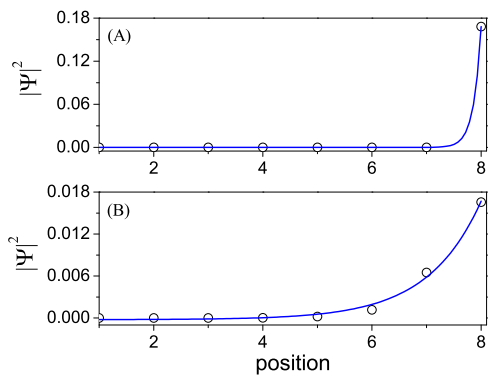}}\caption{(Color
online) The particle density distribution of the edge states: (A): the edge
state of the parent topological superconductor; (B): the edge state of the
mid-gap states induced by the vortex superlattice.}%
\end{figure}

In this part we study the $p_{x}+ip_{y}$ topological SC with a square vortex
superlattice. The lattice constant of the square vortex superlattice is set to
be $d=4a.$ We choose $8a\times4a$ sites to be a unit cell ($8a$ along $x$
direction and $4a$ along $y$ direction). To show the topological properties of
the mid-gap states, we put the system on a cylinder (open boundary condition
along x-direction, periodic boundary condition along y-direction). The results
are shown in Fig.2. From Fig.2, one can see that the energy bands from the
bulk consist of two parts : the energy bands of Bogoliubov quasi-particles and
the mid-bands induced by vortex superlattice. Both energy bands have energy
gaps. Except for the energy bands from the bulk, there exist gapless edge
states. There are two types of edge states: one comes from the parent
topological superconductor due to its nontrivial topological properties, the
other comes from the vortex superlattice. That means the mid-gap states
induced by the vortex superlattice also have nontrivial topological
properties. See the illustration in Fig.2 and Fig.3. Fig.3(A) shows the
particle density distribution of the edge state of the parent topological
superconductor and Fig.3(B) shows the particle density distribution of edge
states of the mid-gap states induced by the vortex superlattice. One can see
that due to a larger energy gap, the localized length into the bulk of the
edges state of the parent topological SC is shorter than that of the edge
states of the mid-gap states.

\subsection{Effective tight-binding Majorana lattice model}

\begin{figure}[ptb]
\centering
\scalebox{0.6}{\includegraphics* [1in,3in][8in,7in]{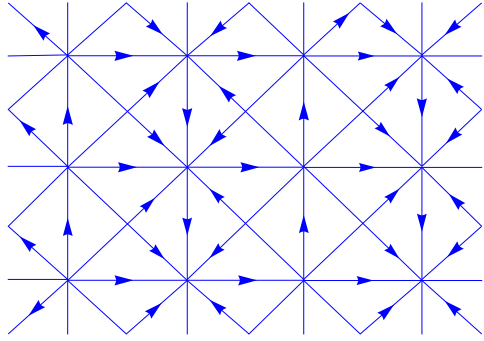}}\caption{(Color
online) The sketch of the square Majorana lattice. The arrow configuration
denotes one possible gauge of the model, $s_{ij}=1$ if arrow from $i$ to $j$,
the clockwise product of the ${s_{ij}}^{\prime}s$ around each triangle is
equal to $-1$ such that each triangular through by $-\pi/2$ flux. And the
gauge choice splits the lattice into two sublattices.}%
\end{figure}

In the last section, we have shown the induced mid-gap energy bands induced by
vortex superlattice. Because each vortex has a residual zero mode and the
inter-vortex tunneling effect leads to the energy splitting of the degenerated
zero modes, we can use a tight-binding Majorana lattice model to describe the
mid-gap states proposed in Ref.\cite{zhou}. The Hamiltonian of this
tight-binding Majorana lattice model can be
\begin{align}
H_{F}  &  =i\sum_{l}\sum_{|i-j|=l_{x}}\frac{t_{l_{x}}}{2}s_{ij}^{l_{x}}%
\gamma_{i}\gamma_{j}+i\sum_{l}\sum_{|i-j|=l_{y}}\frac{t_{l_{y}}}{2}%
s_{ij}^{l_{y}}\gamma_{i}\gamma_{j}\nonumber \\
&  +i\sum_{|i-j|=\sqrt{2}}\frac{t_{\sqrt{2}}}{2}s_{ij}^{\sqrt{2}}\gamma
_{i}\gamma_{j},
\end{align}
where $\gamma_{i}$ is the operator of Majorana fermion in $i_{th}$ vortex core
obeying the self-conjugate condition and the canonical commutate relation $\{
\gamma_{i},\gamma_{j}\}=2\delta_{ij}$. The indices $l=1,2$ denote nearest
neighbor couplings, next-next nearest neighbor couplings, respectively,
$t_{l_{x}},t_{l_{y}}$ are the corresponding coupling strength along $x$ and
$y$ direction, and $t_{\sqrt{2}}$ is the next nearest neighbor coupling
strength. Each $s_{ij}$ connecting bond $\left \langle ij\right \rangle $ has
$Z_{2}$ gauge degree of freedom which does not affect any physical
conclusions. Motivated by the decompose rules that two Majorana fermions fuse
into a complex fermion, and no net flux through the primitive cell, the
condition that there is $-\pi/2$ flux in each triangular plaquette must be
satisfied. Fig.4 shows the Majorana lattice model.

We then decompose Majorana operator as
\begin{equation}
\gamma_{2i-1}=f_{i}+f_{i}^{\dag},\quad i\gamma_{2i}=f_{i}-f_{i}^{\dag},
\end{equation}
where $f_{i}$ destroys a Dirac fermion on the center of link (see Fig.4). This
formalism reproduces an effective spinless SC state which we model as
\begin{equation}
H_{F}=\sum_{\mathbf{k}}\left(
\begin{array}
[c]{cc}%
f_{\mathbf{k}}^{\dag} & f_{-\mathbf{k}}%
\end{array}
\right)  \left(
\begin{array}
[c]{cc}%
\phi_{\mathbf{k}}^{0}+\phi_{\mathbf{k}}^{3} & \phi_{\mathbf{k}}^{1}%
-i\phi_{\mathbf{k}}^{2}\\
\phi_{\mathbf{k}}^{1}+i\phi_{\mathbf{k}}^{2} & \phi_{\mathbf{k}}^{0}%
-\phi_{\mathbf{k}}^{3}%
\end{array}
\right)  \left(
\begin{array}
[c]{c}%
f_{\mathbf{k}}\\
f_{-\mathbf{k}}^{\dagger}%
\end{array}
\right)  .
\end{equation}
The energy dispersion has the form as
\[
E_{k}=\phi_{\mathbf{k}}^{0}\pm|\vec{\phi}_{\mathbf{k}}|,
\]
with $\vec{\phi}_{\mathbf{k}}\equiv(\phi_{\mathbf{k}}^{1},\phi_{\mathbf{k}%
}^{2},\phi_{\mathbf{k}}^{3})$. The vector $\phi_{\mathbf{k}}^{i}$ is expressed
as
\begin{align}
\phi_{\mathbf{k}}^{0} &  =-4t_{2y}\sin2A\mathbf{k}_{y},\nonumber \\
\phi_{\mathbf{k}}^{1} &  =4t_{2x}\sin2A\mathbf{k}_{y}-4t_{1y}\sin
A\mathbf{k}_{y},\nonumber \\
\phi_{\mathbf{k}}^{2} &  =-2t_{1x}\sin2A\mathbf{k}_{x}-4t_{\sqrt{2}}%
\sin2A\mathbf{k}_{x}\cos \mathbf{k}_{y},\nonumber \\
\phi_{\mathbf{k}}^{3} &  =-2t_{1x}\sin2A\mathbf{k}_{x}-4t_{\sqrt{2}}\cos
A\mathbf{k}_{y}+2t_{1x}\nonumber \\
&  -4t_{\sqrt{2}}\cos A\mathbf{k}_{y}\cos2A\mathbf{k}_{x}.
\end{align}
The calculation for these functions is straightforward (though a little
tedious), and the parameter $A=4$ arises from the vortex-distance
$d=4$.\begin{widetext}
\begin{figure}[ptb]
\begin{center}
\centering
\includegraphics[width=1\textwidth]{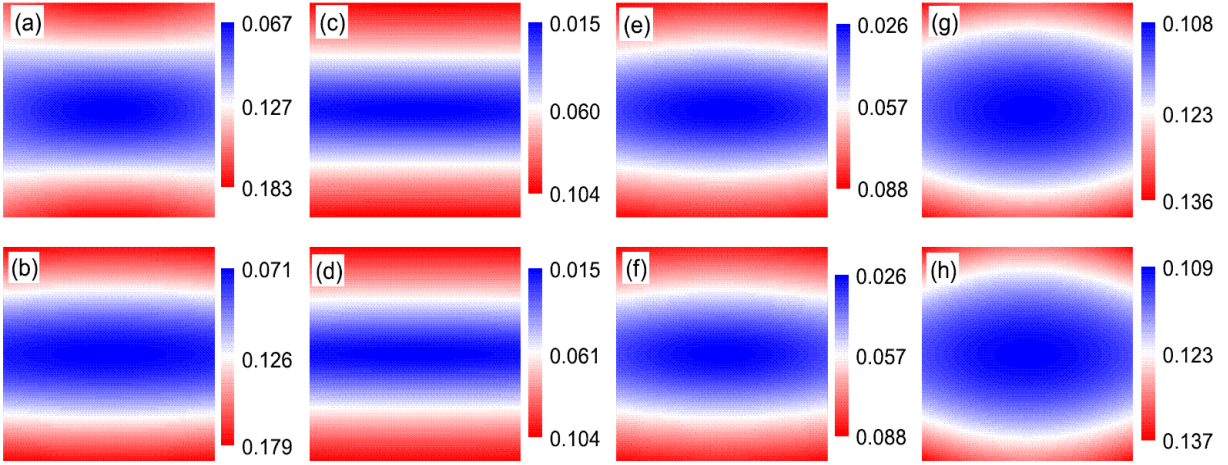}\caption{Top panel: The contour
plot of dispersion of mid-gap energy bands induced by vortex superlattice as a
function of paired order parameter $\Delta$ in the zone $k_{x}\otimes k_{y}%
\in[-0.1,0.1]\otimes[-0.2,0.2]$. The paired order parameter is equal
$0.4,0.8,1.2$, and $1.4$ for $(a),(c),(e),(g)$, respectively.
Bottom panel: The fitting dispersion of the effective tight-binding Majorana
lattice model, and the hopping parameters are obtained in TABLE I.}%
\label{Fig5}%
\end{center}
\end{figure}
\end{widetext}

Next, we obtain the tunneling parameters by fitting the energy dispersion of
the mid-gap states induced by the vortex superlattice. As shown in top panel
of Fig.5, the energy dispersion of the mid-gap states are obtained by the
numerical calculations. The bottom panel shows the energy dispersion of the
effective tight-binding Majorana lattice model and the corresponding hopping
parameters are given in TABLE.1. We point out that the effective tight-binding
Majorana lattice model may describe the mid-gap states very well. So we have
an effective method to calculate the quantum properties of the $p_{x}+ip_{y}$
SC with vortex superlattice quantitatively, and which can be compared with the
experimental research.

\begin{table}[ptb]
\caption{The hopping parameters of the effective Majorana lattice model by
fitting from the mid-gap energy bands induced by vortex superlattice (All the
parameters given under the fixed chemical potential $u=-2.0$). }%
\begin{tabular}
[c]{cccccccc}\hline
& $\Delta=0.4$ \qquad & $\Delta=0.6$ \qquad & $\Delta=0.8$ \qquad &
$\Delta=1.0$ \qquad & $\Delta=1.2$ \qquad & $\Delta=1.4$\qquad & \\ \hline
$t_{1x}$ & 0.02621 & 0.01547 & 0.00741 & 0.01615 & 0.03376 & 0.05682 & \\
$t_{1y}$ & 0.05872 & 0.04294 & 0.03397 & 0.02984 & 0.03904 & 0.05852 & \\
$t_{\sqrt{2}}$ & 0.00897 & 0.00080 & 0.00189 & 0.00325 & 0.01360 & 0.02677 &
\\
$t_{2x}$ & 0.00012 & -0.0032 & -0.00126 & 0.00009 & 0.00168 & 0.00108 & \\
$t_{2y}$ & 0.0 & 0.0 & 0.0 & 0.0 & 0.0 & 0.0 & \\
fitness & 98.72\% & 99.80\% & 99.98\% & 99.99\% & 99.83\% & 95.28\% & \\ \hline
\end{tabular}
\end{table}

\begin{table}[ptb]
\caption{The hopping parameters directly from the energy splitting in Fig.1}%
\begin{tabular}
[c]{cccccccc}\hline
& $\Delta=0.4$ \qquad & $\Delta=0.6$ \qquad & $\Delta=0.8$ \qquad &
$\Delta=1.0$ \qquad & $\Delta=1.2$ \qquad & $\Delta=1.4$\qquad & \\ \hline
$t_{1x}$ & 0.03678 & 0.03646 & 0.05522 & 0.10012 & 0.16906 & 0.25598 & \\
$t_{1y}$ & 0.03679 & 0.03647 & 0.05522 & 0.10017 & 0.16907 & 0.25598 & \\
$t_{\sqrt{2}}$ & 0.00120 & 0.02362 & 0.02596 & 0.02533 & 0.03763 & 0.07586 &
\\
$t_{2x}$ & 0.00854 & 0.00394 & 0.00593 & 0.01573 & 0.04501 & 0.09284 & \\
$t_{2y}$ & 0.00854 & 0.00394 & 0.00593 & 0.01573 & 0.04501 & 0.09284 &
\\ \hline
\end{tabular}
\end{table}

\subsection{Anisotropicity}

An interesting feature of the mid-gap states is the \emph{anisotropy} for the
hopping parameters along x-direction and those along along y-direction. We
define the anisotropic ratio of the effective Majorana lattice model as
\begin{equation}
\alpha=\left \vert \frac{t_{1_{y}}}{t_{1_{x}}}-1\right \vert .
\end{equation}
For the case of $\Delta=0.8t,$ the Majorana lattice model is highly
anisotropic due to a large anisotropic ratio up to $\alpha \simeq3.584$. For
the case of $\Delta=1.4t,$ the anisotropic ratio is smaller which is
about$\ 0.06$. The physical mechanism of the enlarged anisotropic ratio is not
well understand right now and will be explored in the future.

On the other hand, we can also write down an effective tight-binding Majorana
lattice model by calculating the tunneling parameters from the energy
splitting given in Fig.1. The results are given in TABLE.2. One can see that
the effective tight-binding Majorana lattice model from the energy splitting
is almost isotropic, or $\alpha \rightarrow0$. In particular, from Fig.6, one
can see that the tight-binding Majorana lattice model
from the energy splitting always fails to describe the mid-gap states.

\begin{figure}[ptb]
\includegraphics* [width=0.5\textwidth]{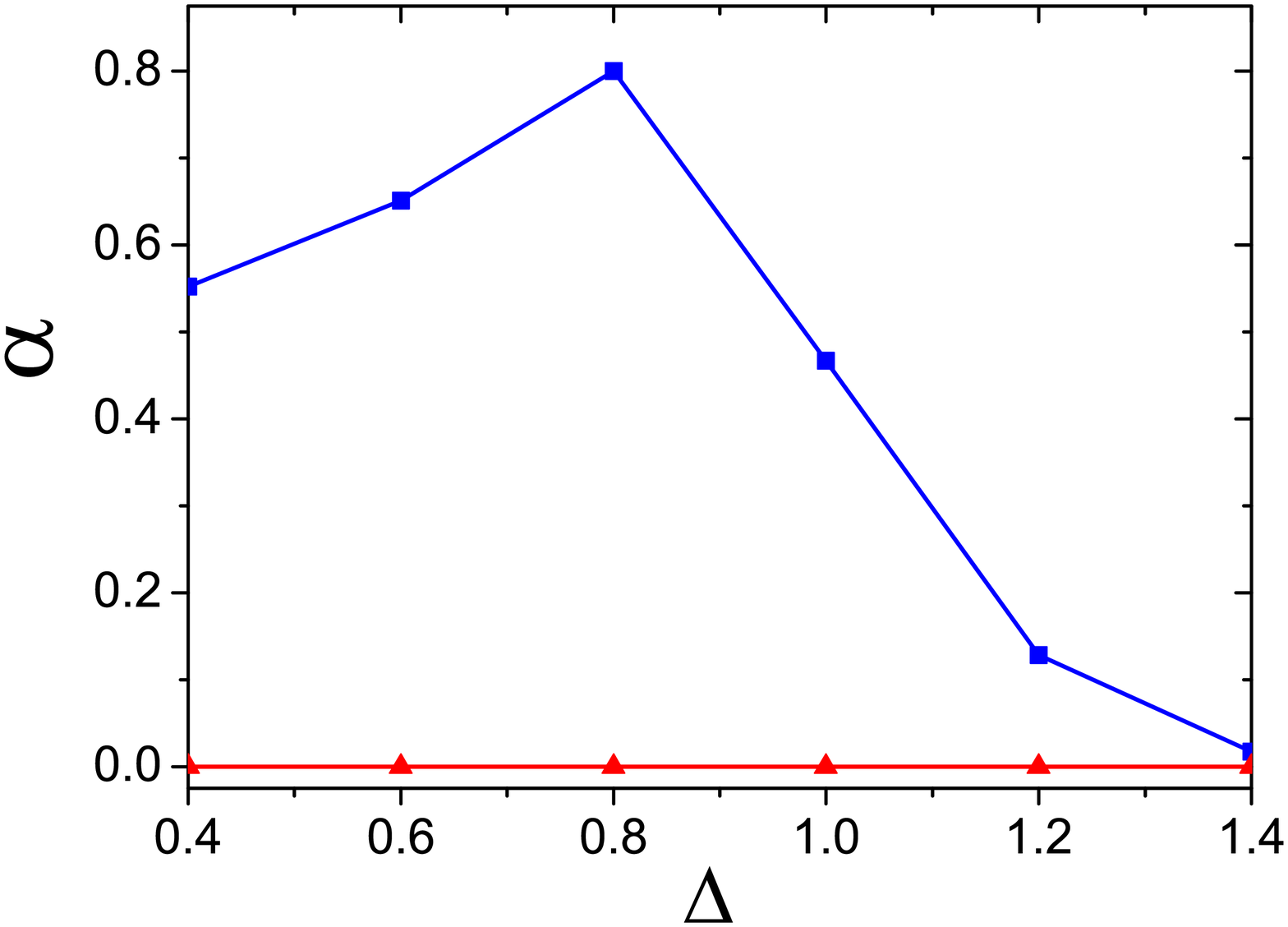}\caption{(Color online) The
anisotropic ratio of the effective Majorana lattice model.}%
\end{figure}

\subsection{Topological invariant}

Another feature of the mid-gap states is the \emph{topology.} We calculate the
topological invariant of the effective tight-binding Majorana lattice model.
The effective tight-binding Majorana lattice model is described by the
Hamiltonian,
\[
H_{F}=\phi_{\mathbf{k}}^{0}\mathcal{I}_{2\times2}+\sum_{i}\sigma_{i}%
\phi_{\mathbf{k}}^{i}.
\]
Thus, the topological invariant of the model is the winding number
\cite{MCheng,XLQi}
\begin{equation}
\mathcal{C}=\int_{\mathbf{k\in BZ}}\frac{d^{2}\mathbf{k}}{4\pi}\vec{\phi}%
\cdot \partial_{\mathbf{k}_{x}}\vec{\phi}\times \partial_{\mathbf{k}_{y}}%
\vec{\phi}, \label{wind}%
\end{equation}
The winding number classifies different homotopy classes mapping from 2D BZ to
2D sphere, $\phi_{n}(\mathbf{k}):T^{2}\rightarrow S^{2}$. Note that the
winding number also associates with the Hamiltonian $\mathcal{H}_{0}=\sum
_{i}\sigma_{i}\phi_{\mathbf{k}}^{i}$. And these two Hamiltonians are
topologically equivalent by a adiabatic deformation connecting $H_{F}$ and
$\mathcal{H}_{0}$ without closed the bulk gap. We perform the integral
Eq.(\ref{wind}) in irreducible Brillouin zone $\mathbf{k}_{x}\otimes
\mathbf{k}_{y}\in \lbrack-\pi/8,\pi/8]\otimes \lbrack-\pi/4,\pi/4]$, which gives
the value $\mathcal{C}=1$ (first column data). So the effective tight-binding
Majorana lattice model is just a "topological SC state" within our
construction formalism. Such hierarchical effect had been employed to
understand the topological quantum transition induced by vortex excitation
that has been studied in the context of Kitaev's honeycomb model supporting
non-Abelian anyon excitation\cite{VLah1,VLah2,zhou}.

\section{Conclusion}

In conclusion, we have studied the $p_{x}+ip_{y}$ topological SC with a square
vortex superlattice. Due to the inter-vortex tunneling, there exist mid-gap
energy bands induced by the vortex superlattice. We found that such mid-gap
energy bands have nontrivial topological properties including the gapless edge
states and non-zero winding number. Then we write down an effective
anisotropic tight-binding Majorana lattice model to characterize the mid-gap
states and obtain the hopping parameters by fitting the energy dispersion with
numerical calculations. In particular, we find that the anisotropic
tight-binding Majorana lattice model has nontrivial topological properties.

\begin{center}
{\textbf{* * *}}
\end{center}

This work is supported by National Basic Research Program of China (973
Program) under the grant No. 2011CB921803, 2012CB921704 and NSFC Grant No.
11174035. \bigskip

\section{{Appendix:} The derivation of BdG equation}

In this appendix, we give the derivation of BdG equation corresponding to
Eq.(\ref{bdg}). The starting point is a general BCS mean field Hamiltonian
\begin{align}
H_{BCS}  &  =\int d\mathbf{r}\psi^{\dag}(\mathbf{r})(-\frac{\partial^{2}}%
{2m}-u)\psi(\mathbf{r})\nonumber \\
&  +\frac{1}{2}\int \int d\mathbf{r}d\mathbf{r}^{\prime}\{ \psi^{\dagger
}(\mathbf{r})\Delta(\mathbf{r},\mathbf{r}^{\prime})\psi^{\dagger
}(\mathbf{r^{\prime}})+h.c\},
\end{align}
where the pairing function $\Delta(\mathbf{r,r}^{\prime})$ is given by
\begin{equation}
\Delta(\mathbf{r,r}^{\prime})=\Delta(\frac{\mathbf{r+r}^{\prime}}{2}%
)(\partial_{x^{\prime}}+i\partial_{y^{\prime}})\delta(\mathbf{r-r}^{\prime}).
\end{equation}

To diagonalize this Hamiltonian, it's useful to perform the Bogoliubov
transformation
\begin{equation}
\psi(\mathbf{r})=\sum_{n}[\gamma_{n}u_{n}(\mathbf{r})+\gamma_{n}^{\dag}%
v_{n}^{\ast}(\mathbf{r})],
\end{equation}
where index $n$ labels $n_{th}$ quasi-particle eigenstates. The complex
functions $u_{n}(\mathbf{r})$ and $v_{n}(\mathbf{r})$ are determined by the
following requirement
\begin{equation}
\lbrack H_{BCS},\text{ }\gamma_{n}]=-E_{n}\gamma_{n}.
\end{equation}
The commutators of the field operator and the Hamiltonian is easily
calculated, which generate
\begin{align}
\lbrack H_{BCS},\text{ }\psi(\mathbf{r})]  &  =-(-\frac{\partial^{2}}%
{2m}-u)\psi(\mathbf{r})\nonumber \\
-\frac{1}{2}[\Delta(\mathbf{r})(\partial_{x}  &  +i\partial_{y})+(\partial
_{x}+i\partial_{y})\Delta(\mathbf{r})]\psi^{\dag}(\mathbf{r}).
\end{align}
Substituting the Bogoliubov transformation $\psi(\mathbf{r})=\sum_{n}%
[\gamma_{n}u_{n}(\mathbf{r})+\gamma_{n}^{\dag}v_{n}^{\ast}(\mathbf{r})]$ into
the commutator and considering the diagonalization condition yield
\begin{align}
\lbrack H_{BCS},\text{ }\psi(\mathbf{r})]  &  =-E_{n}u_{n}(\mathbf{r}%
)\gamma_{n}-E_{n}v_{n}^{\ast}(\mathbf{r})\gamma_{n}^{\dag}\\
&  =-(-\frac{\partial^{2}}{2m}-u)[\gamma_{n}u_{n}(\mathbf{r})+\gamma_{n}%
^{\dag}v_{n}^{\ast}(\mathbf{r})]\nonumber \\
&  =-\frac{1}{2}\{ \Delta(\mathbf{r}),\partial_{x}+i\partial_{y}\}[\gamma
_{n}u_{n}(\mathbf{r})+\gamma_{n}^{\dag}v_{n}^{\ast}(\mathbf{r})].\nonumber
\end{align}

Then the coefficients of operator $\gamma_{n}$ give
\begin{equation}
(-\frac{\mathbf{\partial}^{2}}{2m}-u)u(\mathbf{r})+\frac{1}{2}\{
\Delta(\mathbf{r}),\partial_{x}+i\partial_{y}\}v(\mathbf{r})=Eu(\mathbf{r}).
\end{equation}
And the coefficients of operator $\gamma_{n}^{\dag}$ give (after performing a
conjugation)
\begin{equation}
(\frac{\mathbf{\partial}^{2}}{2m}+u)v(\mathbf{r})-\frac{1}{2}\{ \Delta^{\ast
}(\mathbf{r}),\partial_{x}-i\partial_{y}\}v(\mathbf{r})=Ev(\mathbf{r}).
\end{equation}
These two equations are just the BdG equation
\begin{equation}
H_{BdG}(u(\mathbf{r}),\text{ }v(\mathbf{r}))^{T}=E(u(\mathbf{r}),\text{
}v(\mathbf{r}))^{T}%
\end{equation}
where
\begin{equation}
H_{BdG}=\left(
\begin{array}
[c]{cc}%
-\frac{\partial^{2}}{2m}-u & \frac{1}{2}\{ \Delta(\mathbf{r}),\partial
_{x}+i\partial_{y}\} \\
-\frac{1}{2}\{ \Delta^{\ast}(\mathbf{r}),\partial_{x}-i\partial_{y}\} &
\frac{\partial^{2}}{2m}+u
\end{array}
\right)  .
\end{equation}
It's sufficient to ignore the kinetic term $\frac{-\partial^{2}}{2m}$, and
only consider the one-body part by a varying function $-u(\mathbf{r})$ for
studying the topological properties. It is valid by supposing that the energy
gap is never closed \cite{NRead}.


\begin{thebibliography}{99}                                                                                               %


\bibitem {Wen1}X. G. Wen, Int. J. Mod. Phys. B \textbf{4}, 239 (1990).

\bibitem {Wen2}X. G. Wen, Adv. Phys. \textbf{44}, 405 (1995).

\bibitem {DJThouless}D. J. Thouless, M. Kohmoto, M. P. Nightingale, and M. den
Nijs, Phys. Rve. Lett. \textbf{49}, 405 (1982).

\bibitem {CLKane}C. L. Kane and E. J. Mele, Phys. Rev. Lett. \textbf{95},
226801 (2005).

\bibitem {BABernevig}B. A. Bernevig and S. C. Zhang, Phys. Rev. Lett.
\textbf{96}, 106802 (2006).

\bibitem {NRead}N. Read and D. Green, Phys. Rev. \textbf{B 61}, 10267 (2000).

\bibitem {AdyStern}Ady Stern, Nature, \textbf{464} (2010).

\bibitem {ki2}A. Kitaev, Ann. Phys. \textbf{321}, 2 (2006).

\bibitem {free}M. Freedman, M. Larsen, and Z. Wang, Math. Phys. \textbf{227},
605 (2002).

\bibitem {ge}L. S. Georgiev, Phys. Rev. \textbf{B 74}, 235112 (2006). L. S.
Georgiev, Nucl. Phys. \textbf{B 789}. 552-590 (2008).

\bibitem {CNayak}C. Nayak, S. H. Simon, A. Stern, M. Freedman, and S. Das
Sarma, Rev. Mod. Phys. \textbf{80}, 1083 (2008).

\bibitem {Jalicea}J. Alicea, Rep. Prog. Phys. \textbf{75}, 076501 (2012).

\bibitem {MCheng}M. Cheng, R. M. Lutchyn, V. Galitski, and S. Das Sarma, Phys.
Rev. Lett \textbf{103}, 107001 (2009). M. Cheng, R. M. Lutchyn, V. Galitski,
and S. Das Sarma, Phys. Rev. B \textbf{82}, 094504 (2010).

\bibitem {zhou}J. Zhou, Y. J. Wu, R. L. Wu, S. P. Kou, EPL, \textbf{102} 47005 (2013).

\bibitem {XLQi}X. L. Qi, Y. S. Wu, and S. C. Zhang, Phys. Rev. B \textbf{74},
045125 (2006).

\bibitem {VLah1}V. Lahtinen, A. W. W.Ludwig, J. K. Pachos, and S. Trebst,
Phys. Rev. B \textbf{86}, 075115 (2012).

\bibitem {VLah2}V. Lahtinen, New. J. Phys. \textbf{13}, 075009 (2011).
\end{thebibliography}
\end{document}